\documentstyle[aps,prl,epsf,multicol]{revtex}
\begin{document}
\draft
\title{Equivalent classes of closed three-level systems.}
\author{M. B. Plenio }
\address{Optics Section, The Blackett Laboratory, Imperial College,
London SW7 2BZ, England}

\maketitle
\begin{abstract}
In recent years a significant amount of research in quantum optics
has been devoted to the analysis of atomic three-level systems and
for many physical quantities the same effects have been predicted
for different configurations. These configurations can be split
into essentially two classes. One for which the system contains a
metastable state and another where the system has two close-lying
levels and coherence effects become important. We demonstrate
when and why for a wide range of parameters these two classes are
in fact equivalent for many important physical quantities. A
unified picture underlying a large body of work on these
categories of atomic three-level systems is presented and applied
to some examples.
\end{abstract}


\begin{multicols}{2}
Atomic coherence effects are essential for many important effects
in the response of an atomic system to strong laser irradiation.
The Mollow spectrum \cite{Mollow} for a strongly driven two-level
system has been one of the early results in quantum optics where
atomic coherence plays a significant role. Following the detailed
study of the population dynamics and the spectral response of the
laser-driven two-level system (see e.g. \cite{tls}), theoretical
and experimental interest began to shift towards multi-level
configurations and, in particular, to three-level systems. A large
body of work has been devoted to analyze all of the systems shown
in Figs. 1 and 2. The systems in Figs. 1b and 2b, for example,
exhibit many effects based on quantum coherence such as dark
resonances \cite{dark}, electron shelving \cite{Intermittent2}, 
narrow spectral lines \cite{Narrow2,Narrow2a,paskni98}, electro-magnetically 
induced transparency \cite{kylstra} and lasing without inversion 
\cite{LWIb,Imamoglu}. 
However, many of these, and also other effects, had also been predicted 
for systems such as those in Figs. 1a and 2a, where quantum interference 
does not seem to play a major role. Examples are again electron shelving
\cite{Intermittent1,QJump}, spectral line-narrowing without loss of 
intensity \cite{QJump,Narrow1} electro-magnetically induced transparency
\cite{EITa} and lasing without inversion \cite{LWIa}. Obviously, the 
same effects can be found in apparently different configurations with
or without the use of quantum coherence. This suggests an
underlying structure common to all these systems.

The key result of this letter is a proof that reveals such a common
structure for the two systems depicted in Fig. 1, and an analogous
structure (also based on a partial dressed state picture) 
for the systems in Fig. 2. It is
as a consequence of this common structure, that seemingly {\em different}
systems exhibit for many important quantities the {\em same} physical
behaviour. First steps toward this general result have been found for
special cases and less general level configurations in \cite{equiv1,equiv2}
and in the context of electro-magnetically induced transparency 
for example in \cite{Imamoglu,Fleischhauer}.
We illustrate our results with some examples discussing important
physical quantities such as photon statistics and spectra. These
examples demonstrate, how the common structure developed here can
be used to reveal the common origin of a wide variety of effects
for apparently different system. This structure therefore serves
to unify a large body of work that has been devoted to three-level
systems.

We begin by demonstrating that the systems shown in Fig. 1a and 1b obey
equivalent master equations. The three-level configuration in Fig. 1a
consists of a stable ground state $1$, and the two $i\leftrightarrow 1$
transitions driven by different lasers with Rabi frequencies
$\Omega_{i1}^{(a)}$. The strong

\begin{figure}[hbt]
\epsfxsize=8cm
\epsfbox{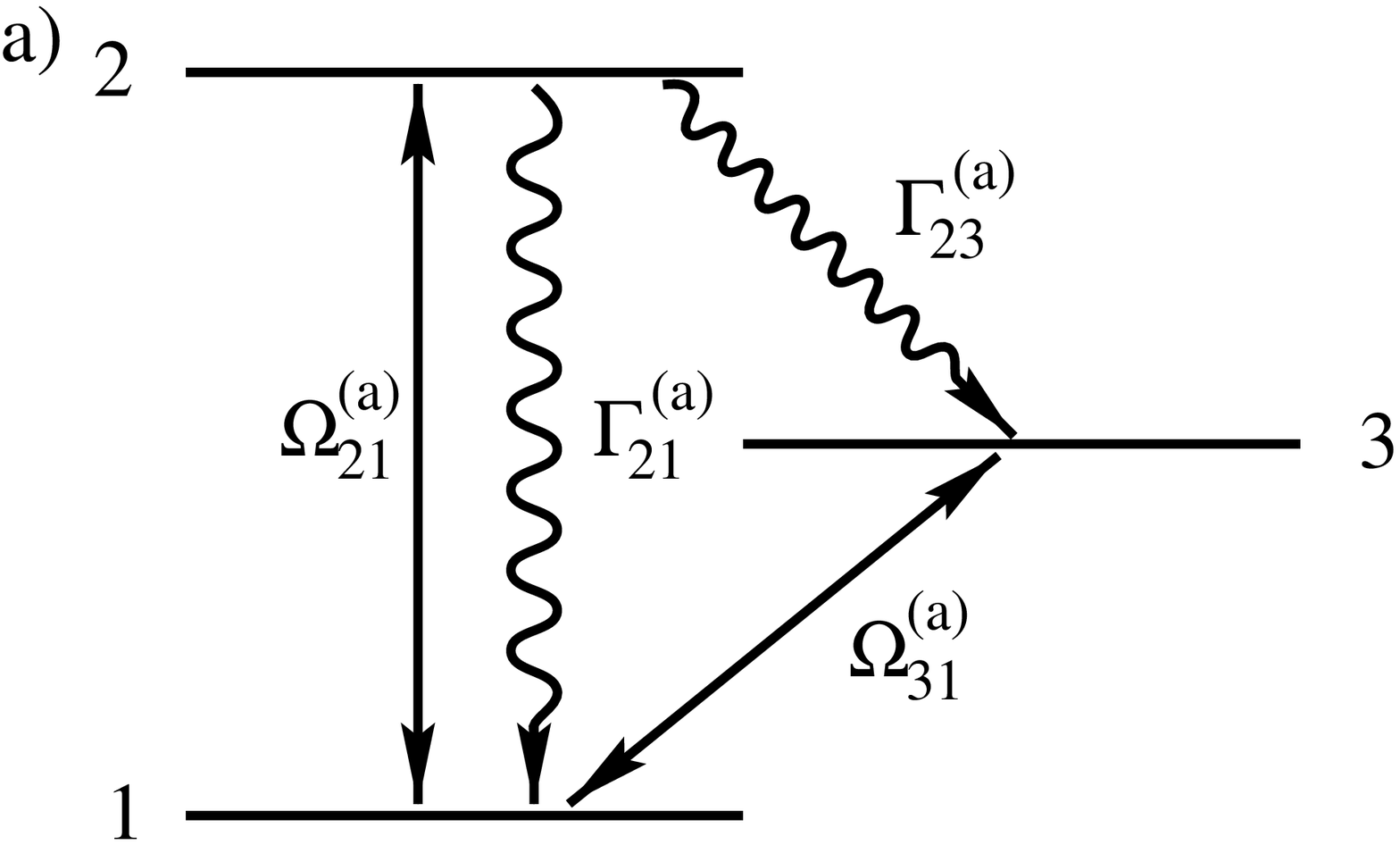}
\vspace*{0.5cm}

\epsfxsize=8cm
\epsfbox{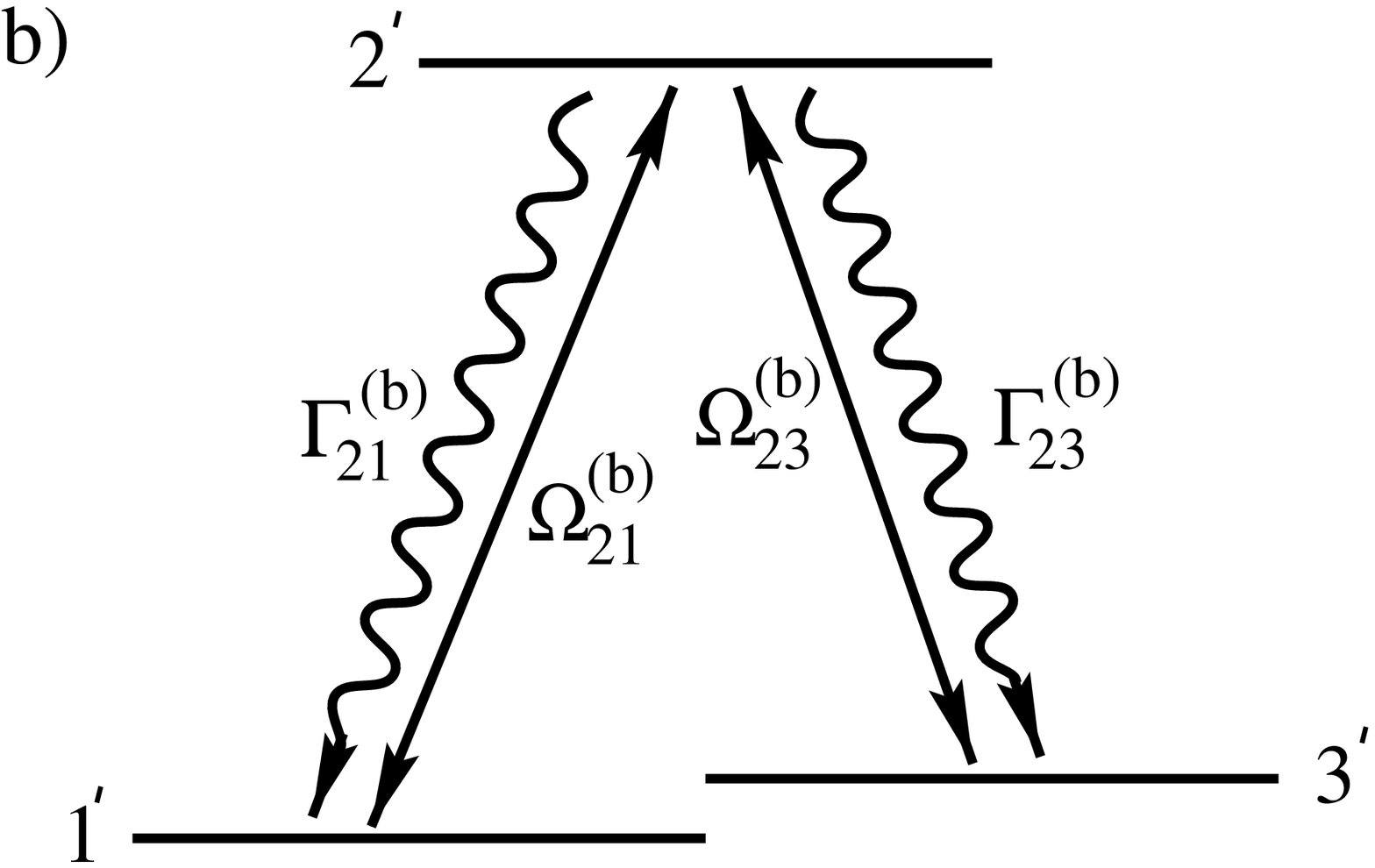}
\vspace*{0.2cm}
\caption{\narrowtext In part a) level $3$ is metastable, while level
$2$ may decay with rates $2\Gamma_{21}^{(a)}$ and $\Gamma_{23}^{(a)}$
to both, level $1$ and level $3$. The $1\leftrightarrow i$ transitions
are driven by {\em two} independent laser fields with Rabi frequencies
$\Omega_{i1}^{(a)}$. In part b) level $2$ is unstable and may decay with
rates $2\Gamma_{12}^{(b)}$ and $\Gamma_{23}^{(b)}$ into the two close-lying
lower levels $1$ and $3$. The system is driven by a {\em single} laser
giving rise to Rabi frequencies $\Omega_{i1}^{(b)}$. The angle between
the dipole moments for the two transitions strongly influences the dynamics
of the system.}
\label{volp}
\end{figure}
\noindent
$1\leftrightarrow 2$ transition decays
at a rate $2\Gamma_{21}^{(a)}$, and the $1\leftrightarrow 3$ transition
is metastable.
The $2\leftrightarrow 3$ transition with spontaneous emission rate
$2\Gamma_{23}^{(a)}$ is assumed to be undriven. For vanishing decay
rate $\Gamma_{23}^{(a)}$ this system is identical to the one proposed
by Dehmelt for the observation of electron shelving \cite{Intermittent1}.
Its photon statistics \cite{Intermittent1,Narrow1}, resonance fluorescence
and absorption spectra \cite{Narrow1} has been analysed in
detail and narrow spectral lines have been found \cite{Narrow1}. Going
over to an interaction picture with respect to
$H_0 = \sum_{i=2}^3 \hbar {\tilde\omega}_{i1} +
    \sum_{{\bf k}\lambda} \hbar\omega_{{\bf k}\lambda}
    a^{\dagger}_{{\bf k}\lambda}a^{{}}_{{\bf k}\lambda}
$ and using standard techniques from quantum optics \cite{Cohen} one can derive
a master equation for the density operator of system 1a. We find
\begin{eqnarray}
    {\dot\rho}^{(a)} &=& \frac{-i}{\hbar}\left[ H_{coh}, \rho^{(a)} \right]
    + 2\Gamma_{21}^{(a)} |1\rangle\langle 2|\rho^{(a)}|2\rangle\langle 1| \nonumber \\
    && + 2\Gamma_{23}^{(a)} |3\rangle\langle 2|\rho^{(a)}|2\rangle\langle 3| \nonumber\\
    && - (\Gamma_{21}^{(a)} + \Gamma_{23}^{(a)}) \left(|2\rangle\langle 2|\rho^{(a)} +\rho^{(a)}|2\rangle\langle 2| \right) \;\; , \label{mastera}
\end{eqnarray}
where $2\Gamma_{ij}^{(a)}=e^2|{\bf d}_{ij}^{(a)}|^2\omega_{ij}^3/(3\pi c^3 \hbar \epsilon_0)$
is equal to the Einstein coefficient on the $i\leftrightarrow j$ transition and
$
    H_{coh} = - \sum_{i=2}^{3} \hbar \Delta_{i1} |i\rangle\langle i|
    + \sum_{i=2}^{3} \hbar\Omega_{i1}^{(a)} \left( |i\rangle\langle 1|
    + |1\rangle\langle i| \right)
$
with the detunings $\Delta_{i1}={\tilde\omega}_{i1} - \omega_{i1}$,
the Rabi frequencies $\Omega_{i1}=-{\bf d}_{i1}^{(a)}{\bf E}/2\hbar$, the
dipole moments ${\bf d}_{ij}^{(a)}$ for the $i\leftrightarrow j$ transition.

Now we would like to show that a basis change leads to the master equations
governing the dynamics of the system shown in Fig. 1b. To see this we introduce
a new basis for which $|2'\rangle = |2\rangle$ and
\begin{equation}
    |1'\rangle = \cos\theta |1\rangle + \sin\theta |3\rangle \;\; ; \;\;
    |3'\rangle = \sin\theta |1\rangle - \cos\theta |3\rangle \label{basis1}
\end{equation}
with
\begin{displaymath}
    \cos\theta = \frac{\Omega_{31}^{(a)}}{\sqrt{\lambda_1^2+(\Omega_{31}^{(a)})^2}}
    \; ; \;
    \lambda_{1} =
    \frac{ -\Delta_3 \pm \sqrt{\Delta_3^2 + 4(\Omega_{31}^{(a)})^2}}{2} .
\end{displaymath}
The basis Eq. (\ref{basis1}) diagonalizes the Hamiltonian
$
    H_{13} = - \hbar \Delta_{31} |3\rangle\langle 3|
    + \hbar\Omega_{31}^{(a)} \left(|3\rangle\langle 1| +
    |1\rangle\langle 3| \right) \;\; ,
$
and it can therefore be viewed as a partial dressed state picture.
Shifting the origin of energy such that it coincides with level $|1'\rangle$
we obtain $H_{coh}$ in the new basis
\begin{eqnarray}
    H'_{coh} &=& - \sum_{i=2}^{3} \hbar {\tilde\Delta}_{i1} |i'\rangle\langle i'|
    -\hbar\Omega_{21}\cos\theta \left( |2'\rangle\langle 1'|
    + |1'\rangle\langle 2'| \right) \nonumber\\
    && -\hbar\Omega_{21}\sin\theta \left( |2'\rangle\langle 3'|
    + |3'\rangle\langle 2'| \right) \;\; ,
\end{eqnarray}
where ${\tilde\Delta}_{21}=\Delta_{21}+\lambda_1$ and
${\tilde\Delta}_{31}=\Delta_{31}+\lambda_1-\lambda_2$.
In the basis given by Eq. (\ref{basis1}) we find the new master equation
\begin{eqnarray}
    {\dot\rho}' &=& \frac{-i}{\hbar}\left[ H'_{coh}, \rho' \right] -
     (\Gamma'_{21}+\Gamma'_{23})(|2'\rangle\langle 2'|\rho' + \rho' |2'\rangle\langle 2'|) \nonumber\\
    && + 2\Gamma'_{21} |1'\rangle\langle 2'|\rho'|2'\rangle\langle 1'|
       + 2\Gamma'_{23} |3'\rangle\langle 2'|\rho'|2'\rangle\langle 3'| \nonumber\\
    && + 2\Gamma'_{13} |1'\rangle\langle 2'|\rho'|2'\rangle\langle 3'|
       + 2\Gamma'_{31} |3'\rangle\langle 2'|\rho'|2'\rangle\langle 1'| \;\; ,
    \label{masternew}
\end{eqnarray}
where $\Gamma'_{21}=\Gamma_{21}^{(a)}\cos^2\theta + \Gamma_{23}^{(a)}\sin^2\theta$,
$\Gamma'_{23}=\Gamma_{21}^{(a)}\sin^2\theta + \Gamma_{23}^{(a)}\cos^2\theta$ and
$\Gamma'_{13}=\Gamma'_{31}=(\Gamma_{21}^{(a)}-\Gamma_{23}^{(a)})\cos\theta\sin\theta$.

The key result is now the observation, that the master equation
(\ref{masternew}) is of the same form as that for the system
shown in Fig. 1b. In fact, if the dipole moments ${\bf d}_{2i}^{(b)}$
for the $2\leftrightarrow i$-transitions in the system in Fig. 1b
form an angle $\phi$, i.e.
$\cos\phi = {\bf d}_{21}^{(b)}{\bf d}_{23}^{(b)}/|{\bf d}_{21}^{(b)}||{\bf d}_{23}^{(b)}|$,
then we find the master equation \cite{equiv1,Cohen}
\begin{eqnarray}
    {\dot\rho}^{(b)} &=& \frac{i}{\hbar} \left[ H_{coh}^{(b)},\rho^{(b)}\right]
     + 2\Gamma_{21}^{(b)} |1'\rangle\langle 2'|\rho^{(b)}|2'\rangle\langle 1'| \nonumber\\
    && + 2\Gamma_{23}^{(b)} |3'\rangle\langle 2'|\rho^{(b)}|2'\rangle\langle 3'| \nonumber\\
    && + 2\sqrt{\Gamma_{23}^{(b)}\Gamma_{21}^{(b)}}\cos\phi
             |3'\rangle\langle 2'|\rho^{(b)}|2'\rangle\langle 1'| \nonumber \\
    && + 2\sqrt{\Gamma_{23}^{(b)}\Gamma_{21}^{(b)}}\cos\phi
         |1'\rangle\langle 2'|\rho^{(b)}|2'\rangle\langle 3'| )\nonumber\\
    && - (\Gamma_{21}^{(b)}+\Gamma_{23}^{(b)}) (|2'\rangle\langle 2'|\rho^{(b)} -
        \rho^{(b)}|2'\rangle\langle 2'| )
    \label{masterb}
\end{eqnarray}
with
$
    H_{coh}^{(b)} = -\hbar\Delta_2|2'\rangle\langle 2'| -
    \hbar\Omega_{21}^{(b)} \left( |2'\rangle\langle 1'|
    + |1'\rangle\langle 2'| \right) 
    + \hbar(\Delta_3-\Delta_2)|3'\rangle\langle 3'|
     -\hbar\Omega_{23}^{(b)} \left( |2'\rangle\langle 3'|
    + |3'\rangle\langle 2'| \right)
$
and $\Delta_2={\tilde\omega_{2}} - \omega_{21}$ and $\Delta_3={\tilde\omega_{3}} - \omega_{31}$.
To see the equivalence between master equations Eq. (\ref{masternew}) and
(\ref{masterb}) we just need to chose $\phi$ such that
\begin{equation}
    \cos^2\phi = \Gamma'_{13}\Gamma'_{31}/\Gamma'_{21}\Gamma'_{23} \;\; ,
    \label{phitheta}
\end{equation}
where the $\Gamma'_{ij}$ have been defined below Eq. (\ref{masternew}).
The polarization of the laser in system 1b is then chosen such
that $\Omega_{21}^{(b)}\sin\theta = \Omega_{23}^{(b)}\cos\theta$ which
can always be satisfied. If in system 1a we chose $\Gamma_{23}^{(a)}=0$,
one can easily verify that the master equation Eq. (\ref{masternew}) is
identical to the master equation Eq. (\ref{masterb}) with $\phi=0$.
Therefore the system in Fig. 1a with $\Gamma_{23}^{(a)}=0$ is equivalent
the system in Fig. 1b with {\em parallel} transition dipole moment!
Furthermore, the master equations for the system in Fig. 1b with
non-parallel dipole moments ($\phi\neq 0$) is equivalent to the master
equation (\ref{masternew}) for the system in Fig. 1a with non-vanishing
$\Gamma_{23}$! Note that the basis change Eq. (\ref{basis1}) implies that
the weakly coupled state $|3\rangle$ in system 1a is a superposition of
the two strongly coupled states in system 1b, i.e. quantum coherence leads
to a metastable superposition state in system 1b.

Following an analogous procedure to the one for the systems in Fig. 1, we
are also able to exhibit the equivalence of the master equations
for the systems shown in Figs. 2a and 2b. In Fig. 2a
both the $2\leftrightarrow 1$ and the $2\leftrightarrow 3$ transition are
driven by individual lasers at Rabi frequencies $\Omega_{21}^{(a)}$
and $\Omega_{23}^{(a)}$ respectively. The $2\leftrightarrow 3$ transition
is assumed to be metastable while the other two transitions decay with rates
$2\Gamma_{21}^{(a)}$ and $2\Gamma_{23}^{(b)}$ respectively. The system in
Fig. 2b is a V-system where the two upper levels $2$ and $3$ are both unstable
and can decay with rates $2\Gamma_{21}^{(b)}$ and $2\Gamma_{31}^{(b)}$ to the
common ground state $1$. The system is driven by a single laser which gives
rise to Rabi frequencies $\Omega_{21}^{(b)}$ and $\Omega_{31}^{(b)}$.
The dynamics of the system 2b depends strongly on the relative orientation
of the dipole moments ${\bf d}_{i1}^{(b)}$ on the $i\leftrightarrow 1$
transition. For the system in Fig. 2a we find the master equation
\begin{eqnarray}
    {\dot\rho}^{(a)} &=& -\frac{i}{\hbar} [H_{coh},\rho^{(a)}]
    + 2 \sum_{i=2}^{3} \Gamma_{i1}^{(a)} |1\rangle\langle i|\rho^{(a)}|i \rangle\langle 1|
    \nonumber\\
    && - \sum_{i=2}^{3} \Gamma_{i1}^{(a)}
    \left( |i\rangle\langle i|\rho^{(a)} + \rho^{(a)} |i\rangle\langle i| \right) \label{master2a}
    \;\; ,
\end{eqnarray}
where
$H_{coh} = -\hbar\Delta_2 |2\rangle\langle 2| +
    \hbar\Omega_{21}^{(a)} \left(|1\rangle\langle 2| + |2\rangle\langle 1| \right)
    + \hbar(\Delta_3-\Delta_2)|3\rangle\langle 3|  +
    \hbar\Omega_{23}^{(a)}\left(|3\rangle\langle 2| + |2\rangle\langle 3| \right).$
with $\Delta_2={\tilde\omega_{2}} - \omega_{21}$ and $\Delta_3={\tilde\omega_{3}} - \omega_{23}$.
Now consider the basis change for which $|1'\rangle = |1\rangle$ and
\begin{equation}
    |2'\rangle = \cos\theta |2\rangle + \sin\theta |3\rangle \; ; \;
    |3'\rangle = \sin\theta |2\rangle - \cos\theta |3\rangle \label{basis2} .
\end{equation}
with
\begin{eqnarray}
    \cos\theta &=& \frac{ \Omega_{23}^{(a)} }{\sqrt{\lambda_1^2 + (\Omega_{23}^{(a)})^2}}\\
    \lambda_{1/2} &=& \frac{1}{2}\left(\Delta_3 \pm
    \sqrt{\Delta_3^2 + 4(\Omega_{23}^{(a)})^2} \right) - \Delta_2 \; .
\end{eqnarray}
Using the basis change Eq. (\ref{basis2}) we obtain with
$
    H'_{coh} = \hbar\lambda_1 |2'\rangle\langle 2'| + \hbar\Omega_{21}\cos\theta
    ( |2'\rangle\langle 1'| + |1'\rangle\langle 2'| ) 
    +\hbar\lambda_2|3'\rangle\langle 3'| + \hbar\Omega_{21}\sin\theta
    ( |3'\rangle\langle 1'| + |1'\rangle\langle 3'| )
$
the new Bloch equations
\begin{eqnarray}
    {\dot\rho}' &=& \frac{-i}{\hbar} [ H_{coh},\rho']
       + \sum_{i=2}^{3} 2\Gamma'_{i1}|1'\rangle\langle i'|\rho'|i'\rangle\langle 1'|\nonumber\\
    &-& \sum_{i=2}^{3} \Gamma'_{i1} (|i'\rangle\langle i'|\rho' + \rho'|i'\rangle\langle i'|) \nonumber\\
    &-& \Gamma'_{32} (|3'\rangle\langle 2'|\rho' + \rho'|2'\rangle\langle 3'|)
       - \Gamma'_{23} (|2'\rangle\langle 3'|\rho' + \rho'|3'\rangle\langle 2'|) \nonumber\\
    &+&  2 \Gamma'_{23}|1\rangle\langle 2'|\rho'|3'\rangle\langle 1'|
       + 2\Gamma'_{32}|1\rangle\langle 3'|\rho'|2'\rangle\langle 1'| \; .
    \label{master2b}
\end{eqnarray}
Here $\Gamma'_{21}=\Gamma_{21}\cos^2\theta + \Gamma_{31}\sin^2\theta$,
$\Gamma'_{31}=\Gamma_{21}\sin^2\theta + \Gamma_{31}\cos^2\theta$ and
$\Gamma'_{32}=\Gamma'_{23}=(\Gamma_{21}-\Gamma_{31})\cos\theta\sin\theta$.

If the angle between the dipole moments ${\bf d}_{i1}^{(b)}$ on the
$i\leftrightarrow 1$-transition in system 2b equals $\phi$,
we use the correspondence between $\phi$ and $\theta$
\begin{equation}
    \cos^2\phi = \Gamma'_{32}\Gamma'_{23}/
    \Gamma'_{21}\Gamma'_{31} \;\; . \label{corr2}
\end{equation}
and chose the polarization ${\bf E}$ of the laser such that
${\bf d}_{2'1'}^{(b)}{\bf E}=2\hbar\Omega_{21}\cos\theta$ and
${\bf d}_{3'1'}^{(b)}{\bf E}=2\hbar\Omega_{21}\sin\theta$. Then
we find that the master equation Eq. (\ref{master2b}) is exactly
equivalent to the Bloch equations for the V-system with close-lying
upper levels given in Fig. 2b.
\begin{figure}[htb]
\epsfxsize=8cm
\epsfbox{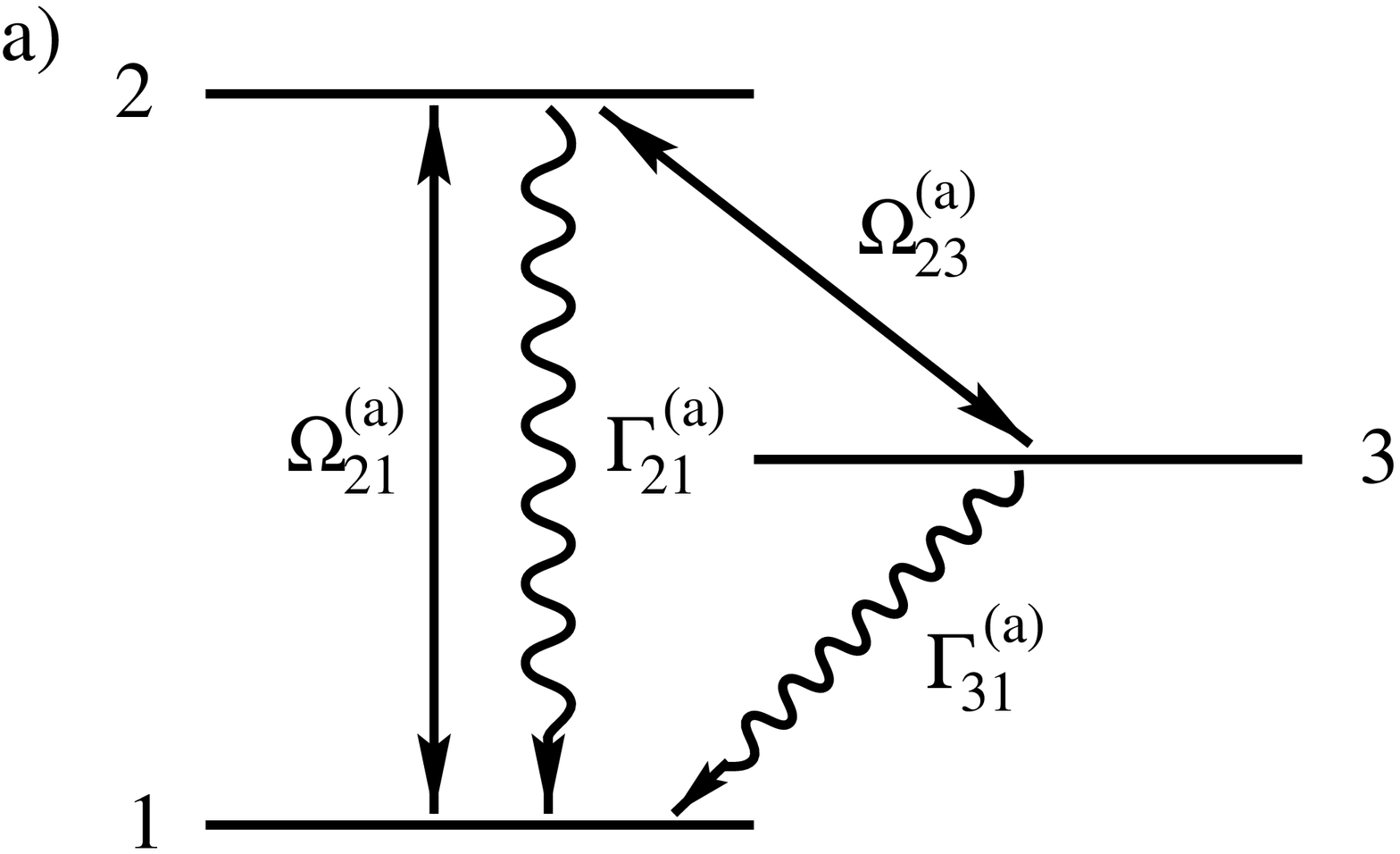}
\vspace*{0.5cm}

\epsfxsize=8cm
\epsfbox{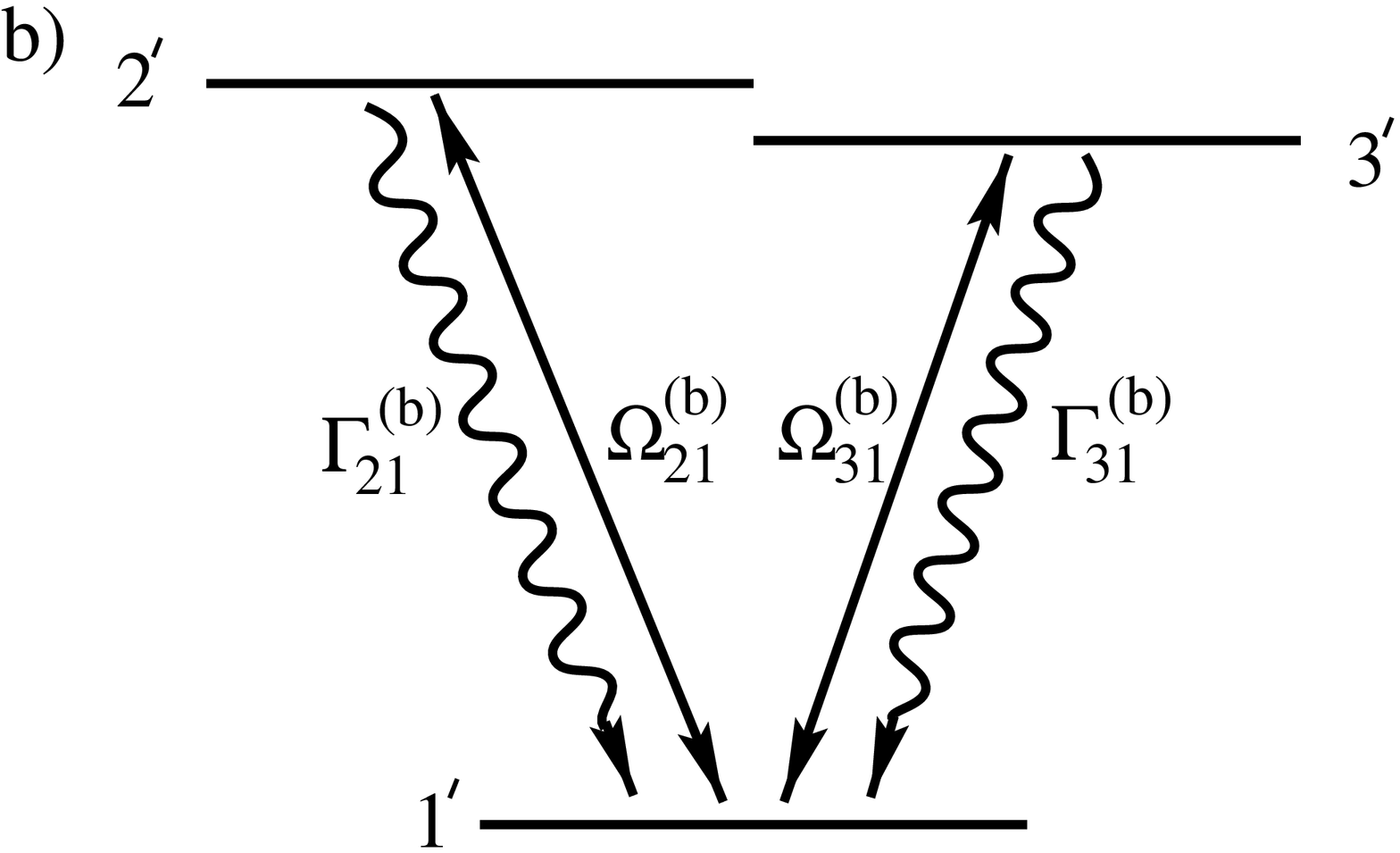}
\vspace*{0.2cm}
\caption{\narrowtext In part a) the $2\leftrightarrow 3$ transition
is metastable, while the $i\leftrightarrow 1$ transitions decay with
rates $\Gamma_{i1}^{(a)}$. The $2\leftrightarrow 1$
and $2\leftrightarrow 3$ transition are driven by {\em two} independent
laser fields with Rabi frequencies $\Omega_{21}^{(a)}$ and $\Omega_{23}^{(a)}$.
In part b) both levels $2$ and $3$ are lying very close, and may
decay with rates $2\Gamma_{i1}^{(b)}$ into level $1$. The system is
driven by a {\em single} laser driving both transitions. The angle
between the dipole moments for the two transitions strongly influences the
dynamics of the system.}
\label{lovp}
\end{figure}
To demonstrate that the basis changes introduced above are useful,
we will now show for some relevant physical quantities that the two
systems in Fig. 2 can exhibit exactly the same behaviour. We begin
with the intensity correlation function of the light emitted from
the atoms. For system 2a the intensity correlation function is given
by
\begin{equation}
    g^{(2)}_{a}(\tau) = 2\langle \Gamma_{21}^{(a)}\sigma_{22}(\tau) +
    \Gamma_{31}^{(a)}\sigma_{33}(\tau)\rangle \label{co1}
\end{equation}
where the average is taken for a system initially in the ground state
$|1\rangle$. With Eq. (\ref{basis2}) and (\ref{corr2})
we find
\begin{eqnarray}
    g^{(2)}_{a}(\tau) &=& 2\langle \Gamma'_{21}\sigma_{2'2'}(\tau) +
    \Gamma'_{31}\sigma_{3'3'}(\tau) \nonumber \\
    && \;\;\;\; +
    \sqrt{\Gamma'_{21} \Gamma'_{31} }\cos\phi
    (\sigma_{2'3'}(\tau) + \sigma_{3'2'}(\tau) ) \rangle \; \label{co2}
\end{eqnarray}
where the average is again taken in the ground state $|1'\rangle$.
Eq. (\ref{co2}) is identical to the intensity correlation function
$g^{(2)}_{b}(\tau)$ of the system in Fig. 2b if we chose $\Gamma_{ij}^{(b)}=\Gamma'_{ij}$
\cite{Plenio5}. It should be noted that identical intensity correlation
functions for the two systems 2a and 2b imply that also the
{\em next} photon probabilities for the two systems coincide
and therefore all properties of the photon statistics \cite{QJump}.

Now consider the spectrum of the resonance fluorescence of the two systems.
For the spectrum of system 2a we find with Eq. (\ref{basis2})
\begin{eqnarray}
    S(\omega) &\sim& \Re\! \int_{0}^{\infty} \!\!\!\!\!\! d\tau e^{-i\omega\tau}
    \langle\sigma_{21}(\tau)\sigma_{12}\rangle_{ss} \nonumber\\
    &=& \Re\! \int_{0}^{\infty} \!\!\!\!\!\! d\tau e^{-i\omega\tau} \langle
    \sigma_{2'1'}(\tau)\sigma_{1'2'}\cos^2\theta +
    \sigma_{3'1'}(\tau)\sigma_{1'3'}\sin^2\theta  \nonumber\\
    &&+(\sigma_{2'1'}(\tau)\sigma_{1'3'} + \sigma_{3'1'}(\tau)\sigma_{1'2'})\cos\theta\sin\theta
    \rangle_{ss}  . \label{spectrum}
\end{eqnarray}
To maximize the visibility of coherence effects in the spectrum we
assume that we observe the resonance fluorescence spectrum of system 2b
along the polarization direction $\hat{\bf E}$ of the laser.
If we chose $\hat{\bf E}$ such that
$\hat{\bf E}{\bf d}_{21}^{(b)}/\hat{\bf E}{\bf d}_{31}^{(b)}=\cos\theta/\sin\theta$
the spectrum for system 2b coincides with Eq. (\ref{spectrum}).
Analogously we can show that the absorption spectra for the two systems
can be made to coincide.

As an application of the above results, we demonstrate that a number
of recent results, which have previously been considered as due to
different mechanisms, are in fact essentially equivalent. It has
been pointed out that the system in Fig. 2a can exhibit bright and dark
periods in the resonance fluorescence if the metastable $2\leftrightarrow 3$
transition is weakly driven \cite{Intermittent1,Narrow1}. Quantitatively
the same behaviour has been predicted later for the system in Fig 2b
\cite{Intermittent2}. The fact that the {\em next} photon probabilities
for the two systems can be made to coincide (see discussion below Eq.
(\ref{co2})), clarifies why electron shelving is possible in both systems.
In system 2a the electron is 'shelved' in the metastable state $|3\rangle$,
while in system 2b the electron is 'shelved' in a weakly coupling
coherent superposition of the two upper levels.

We are also able to unify recent results on the resonance fluorescence and
absorption spectra for different three-level configurations using our approach.
In \cite{Narrow1} it has been shown analytically, that in system 2a one can
observe an ultra-narrow peak in the spectrum of resonance fluorescence
in the same parameter regime in which the system exhibits electron shelving. 
This peak can be understood quantitatively as a widening of the coherent Rayleigh peak due
to the on-off modulation of the light intensity by the electron shelving \cite{Narrow1}.
Subsequently it has been proposed that the system presented in Fig. 2b exhibits
the same effect \cite{Narrow2}. There the angle between the dipole moments on the two
transitions was used to adjust the width of a narrow peak in the spectrum of
resonance fluorescence. The reason for the existence of the narrow peak in system
2b can now be understood quantitatively by considering the dynamics of system 2a for
different decay rates $\Gamma_{31}^{(a)}$ on the $3\leftrightarrow 1$ transition
(see Eq. (\ref{corr2})). As the value of $\Gamma_{31}^{(a)}$ is increased, the
electron can escape the shelving state quicker. Therefore the frequency of modulation
of the resonance fluorescence increases and consequently the width of the narrow peak.
Quantitative agreement for the behaviour between the two systems can be reached with
the approach described above. Following an identical argument, analogous equivalences 
can be demonstrated in the absorption spectrum of the systems shown in the Figures 1 and 2. 
Similar equivalences are known in the context of lasing without inversion and 
electro-magnetically induced transparency \cite{Imamoglu,Fleischhauer}.

These examples show, that a large body of work on the different
three-level systems can be understood and explained quantitatively
analyzing just one system. The common underlying structure of the different systems explains
why and when the same physical effects can be found in different systems.
It is the hope that these results will help to focus future efforts in
theoretical analysis of these systems and will help experimental studies
of proposed effects as it allows the use of alternative systems.

The author thanks S.F. Huelga and P.L. Knight for reading the manuscript 
and the UK Engineering and Physical Sciences Research Council, the 
European Union and the Leverhulme Trust for financial support.

\end{multicols}
\end{document}